\documentclass[11pt]{article}

\usepackage{amsmath,a4wide,array,theorem}

\renewcommand{\baselinestretch}{1.1}

\theoremstyle{break}\theorembodyfont{\rmfamily}\newtheorem{Alg}{Algorithm}

\begin{document}

\newcommand{\Pol}{{\bf P}}
\newcommand{\np}{n}
\newcommand{\xim}{\xi_{\textrm{m}}}
\newcommand{\df}{=}
\newcommand{\lar}{\leftarrow}
\newcommand{\ra}{\rightarrow}
\newcommand{\rambo}{\texttt{RAMBO}}
\newcommand{\sarge}{\texttt{SARGE}}
\newcommand{\Aqcd}{A^{\textrm{QCD}}}
\newcommand{\eqn}[1]{Eq.~\!(\ref{#1})}
\newcommand{\fig}[1]{Fig.~\!\ref{#1}}
\newcommand{\ip}[2]{(#1 #2)}
\newcommand{\scm}{s}
\newcommand{\sqs}{\sqrt{\scm}}
\newcommand{\epl}{e^+}
\newcommand{\emi}{e^-}
\newcommand{\q}{q}
\newcommand{\qb}{\bar{q}}
\newcommand{\gl}{g}
\newcommand{\GeV}{\textrm{GeV}}
\newcommand{\Nge}{N_{\textrm{ge}}}
\newcommand{\Nac}{N_{\textrm{ac}}}
\newcommand{\tcpu}{t_{\textrm{cpu}}}
\newcommand{\epu}[1]{{10}^{#1}}
\newcommand{\emu}[1]{\hspace{1pt}\textrm{e\hspace{1pt}-}\hspace{1pt}#1}
\newcommand{\Hour}{\hspace{1pt}\textrm{h}}

\title{{\bf \texttt{SARGE}: an algorithm for generating QCD-antennas}}


\author{Andr\'e van Hameren\thanks{andrevh@sci.kun.nl, 
                                   $^\dagger$kleiss@sci.kun.nl,
				   $^\ddagger$petros@sci.kun.nl} ~and 
        Ronald Kleiss$^\dagger$\\
	University of Nijmegen, Nijmegen, the Netherlands\\
	\and
	Petros D. Draggiotis$^\ddagger$\\
	University of Nijmegen, Nijmegen, the Netherlands\\
	Institute of Nuclear Physics, NCSR 
$\Delta\eta\mu\acute{o}\kappa\varrho\iota\tau o\varsigma$, Athens, Greece}

\maketitle

\begin{abstract}
We present an algorithm to generate any number of random massless momenta in
phase space, with a distribution that contains the kinematical pole structure
that is typically found in multi-parton QCD-processes. As an application, we
calculate the cross-section of some $\epl\emi\to\textrm{partons}$ processes, 
and compare \sarge's performance with that of the uniform-phase space generator 
\rambo.
\end{abstract}

\noindent
Considering that many multi-jet processes will occur in future hadron colliders,
such as the LHC, it is necessary to calculate their cross-sections. A part of
the amplitude of these processes consists of a multi-parton QCD-amplitude, and
it is well known \cite{Kuijf} that the leading kinematic singularity structure
of the squared matrix elements is given by the so-called {\em antenna pole
structure} (APS). In particular, for $n$ gluons it is given by all permutations
in the momenta of 
\begin{equation}
   \frac{1}{\ip{p_1}{p_2}\ip{p_2}{p_3}\ip{p_3}{p_4}
            \cdots\ip{p_{n-1}}{p_{n}}\ip{p_{n}}{p_1}} \;\;,
\end{equation}
where $\ip{p_i}{p_j}$ denotes the Lorentz invariant scalar product of the gluon
momenta $p_i$ and $p_j$. Actually, it is this kinematical structure that is
implemented in algorithms based on the so called {\texttt{SPHEL}
approximation to calculate the amplitudes \cite{Kuijf}. But it is expected, and
observed, that the same structure occurs in the exact matrix elements
\cite{DKP,CMMP}.

For the integration of the differential cross-sections of the processes under
consideration, the Monte Carlo method is the only option, and a phase space
generator is needed. \rambo\ \cite{SKE} is a robust and efficient algorithm to
generate any number of random massless momenta in their center-of-mass frame
(CMF) with a given energy. However, \rambo\ generates the momenta distributed
uniformly in phase space, so that a large number of events is needed to
integrate integrands with the APS to acceptable precision.  Especially when the
evaluation of the integrand is time-consuming, which is the case for the exact
matrix elements, this is highly inconvenient. 

In this paper, we introduce \sarge, an algorithm to generate any number of
random massless momenta in their CMF with a given energy, distributed with a
density that contains the APS. We shall show that it takes account for a
substantial reduction in computing time in the calculation of cross-sections of
multi-parton processes. We briefly sketch the outline of the \sarge-algorithm;
a fuller discussion, appropriate to hadronic initial states as well, will be 
given elsewhere \cite{DHK2}.

The name \sarge\ stands for {\tt S}taggered {\tt A}ntenna
{\tt R}adiation {\tt GE}nerator, and is inspired by the structure of the
algorithm. It consists of the repeated use of the {\it basic antenna} density
for the generation of a momentum $k$, given two momenta $p_1$ and $p_2$:
\begin{equation}
   dA(p_1,p_2;k) 
   \;\df\; d^4k\,\delta(k^2)\,\theta(k^0)\,
           \frac{1}{\pi}\,\frac{\ip{p_1}{p_2}}{\ip{p_1}{k}\ip{k}{p_2}}\;
           g\left(\frac{\ip{p_1}{k}}{\ip{p_1}{p_2}}\right)
           g\left(\frac{\ip{k}{p_2}}{\ip{p_1}{p_2}}\right)\;\;.
\label{Eq001}
\end{equation}
Here, $g$ is a function that serves to regularize the infrared and collinear
singularities, as well as to ensure normalization over the whole space for $k$:
therefore, $g(\xi)$ has to vanish sufficiently fast for both $\xi\to0$ and
$\xi\to\infty$. 
At this point, we take the simplest possible function we can
think of, that has a sufficiently regularizing behavior. We introduce a
positive non-zero number $\xim$ and take
\begin{equation}
   g(\xi) \;\df\; \frac{1}{2\log\xim}\,\theta(\xi-\xim^{-1})
                                       \theta(\xim-\xi) \;\;,
\label{Eq002}   
\end{equation}
which forces the value of $\xi$ to be between $\xim^{-1}$ and $\xim$, and is 
normalized such that $\int dA=1$. 
Let us immediately adopt the notation 
\begin{equation}
   \xi_1 \df \frac{\ip{p_1}{k}}{\ip{p_1}{p_2}} \quad\quad\textrm{and}\quad\quad
   \xi_2 \df \frac{\ip{k}{p_2}}{\ip{p_1}{p_2}} \;\;.
\end{equation}
The main motivation to make the regularizing function depend on $\xi_1$ and 
$\xi_2$ is that it makes $dA$ completely invariant under Lorentz-and scale 
transformations of the momenta. Consequently, 
the number $\xim$ gives a cut-off for the quotients $\xi_1$ and $\xi_2$ of 
the scalar products of the momenta, and not for the scalar products themselves.
It is, however, possible to relate $\xim$ to the total energy $\sqs$ in the CMF 
and a cut-off $s_0$ on the invariant masses, i.e., the requirement that 
\begin{equation}
   (p_i+p_j)^2 \geq s_0
\label{Eq007}   
\end{equation}
for all pairs of momenta $p_i\neq p_j$. This can be done by choosing 
\begin{equation}
   \xim \;\df\; \frac{\scm}{s_0} - \frac{(\np+1)(\np-2)}{2} \;\;,
\label{Eq003}   
\end{equation}
where $\np$ is the total number of momenta. With this choice, the invariant
masses $(p_1+k)^2$ and $(k+p_2)^2$ are regularized, but can still be
smaller than $s_0$ so that the whole of the demanded phase space is covered.
The $s_0$ can be derived from physical cuts $p_T$ on the transverse momenta and
$\theta_0$ on the angles between the outgoing momenta:
\begin{equation}
   s_0 \;=\;
   2p_T^2\cdot\min\bigg(1-\cos\theta_0\,,\,
                        \left(1+\sqrt{1-p_T^2/\scm}\right)^{-1}\bigg) \;\;.
\end{equation}
We now give the algorithm to generate $k$ under the basic antenna density.
Let $k^0$, $\phi$ and $\theta$ denote the absolute value, the polar angel
and the azimuthal angle of $\vec{k}$ in the frame for which
$\vec{p}_1=-\vec{p}_2$ with $\vec{p}_1$ along the positive $z$-axis. 
To generate $k$, one should 
\begin{Alg}[\texttt{BASIC ANTENNA}]
\begin{enumerate}
\item determine the direction of $\vec{p}_1$ in the CMF of $p_1$ and $p_2$;
\item generate two numbers $\xi_{1}$, $\xi_{2}$ independently, each from the
      density $g(\xi)/\xi$;
\item compute from these the values $k^0$ and $\cos\theta$;
\item generate $\phi$ uniformly in $[0,2\pi)$;
\item construct the momentum $k$ in the CMF of $p_1$ and $p_2$;
\item boost the result to the actual frame in which $p_1$ and $p_2$
 were given.
\end{enumerate}
\end{Alg}

The \rambo\ algorithm was developed with the aim to generate the flat phase
space distribution of $\np$ massless momenta as uniformly as possible. The differential density is given by 
\begin{equation}
  dV_\np(\{p\}) 
  \;\df\; \delta(\sqs-P^0)\delta^3(\vec{P}\,)
        \prod_{i=1}^\np d^4p_i\,\delta(p_i^2)\,\theta(p_i^0) \;\;,
\end{equation}
where $P \df \sum_{i=1}^{\np}p_i$. 
Let us denote 
\begin{equation}
   dA^i_{j,k}\df dA(q_j,q_k;q_i) \;\;,\qquad\textrm{and}\qquad
   \xi^{i,j}_{k,l}\df\frac{\ip{p_i}{p_j}}{\ip{p_k}{p_l}}  \;\;.
\end{equation}
To include the APS in the density, one should
\begin{Alg}[\texttt{QCD ANTENNA}]
\begin{enumerate}
\item generate massless momenta $q_1$ and $q_{\np}$ in CMF;
\item generate $n-2$ momenta $q_j$  by the basic
      antennas $dA^2_{1,{\np}}dA^3_{2,{\np}}dA^4_{3,{\np}}\cdots 
                dA^{{\np}-1}_{{\np}-2,{\np}}$;
\item compute $Q = \sum_{j=1}^{\np}q_j$, and the
      boost and scaling transforms that bring $Q^0$ to $\sqs$\\ 
      and $\vec{Q}$ to $(0,0,0)$;
\item for $j=1,\ldots,{\np}$, boost and scale the $q_j$ accordingly, into the 
      $p_j$.
\end{enumerate}
\end{Alg}
This way, the momenta $p_j$ are generated with differential density
$dV_{\np}(\{p\})\Aqcd_{\np}(\{p\})$, where
\begin{align} 
   \Aqcd_{\np}(\{p\}) 
   \;=\; \frac{\scm^2}{2\pi^{n-1}}\cdot
         \frac{g(\xi^{1,2}_{1,{\np}})g(\xi^{2,{\np}}_{1,{\np}})
                      g(\xi^{2,3}_{2,{\np}})g(\xi^{3,{\np}}_{2,{\np}})\cdots
                      g(\xi^{{\np}-2,{\np}-1}_{{\np}-2,{\np}})
		      g(\xi^{{\np}-1,{\np}}_{{\np}-2,{\np}})}
	 {\ip{p_1}{p_2}\ip{p_2}{p_3}\ip{p_3}{p_4}\cdots
                   \ip{p_{\np-1}}{p_{\np}}\ip{p_{\np}}{p_1}} \;\;. 
\label{Eq005}		      
\end{align}
We point out that, whereas the product $dA^2_{1,{\np}}\cdots
dA^{\np-1}_{\np-2,\np}$ contains a factor $\ip{p_1}{p_{\np}}$ in the
numerator, the scaling transformation carries a Jacobian that is precisely
$\scm^2/\ip{p_1}{p_{\np}}^2$, thus leading to a perfectly symmetric APS.

Usually, the event generator is used to generate cut phase space. 
If a generated event does not satisfy the physical cuts, it is rejected. In the
calculation of the weight coming with an event, the only contribution coming
from the functions $g$ is, therefore, their normalization. In total, this gives
a factor $1/(2\log\xim)^{2{\np}-4}$ in the density.

Because we are dealing with gluon momenta, we want to symmetrize the density. 
This can be done by re-labeling the momenta using a random permutation:
\begin{Alg}[\texttt{SYMMETRIZATION}]
\begin{enumerate}
\item generate a random permutation $\sigma\in S_{\np}$ and put
      $p_i\leftarrow p_{\sigma(i)}$ for all $i=1,\ldots,\np$. 
\end{enumerate}
\end{Alg}
An algorithm to generate the random permutations can be found in \cite{Knuth}.
As a result, the differential density becomes
\begin{align}
   dV_{\np}(\{p\})\left(\frac{1}{\np!}\sum_{\textrm{perm.}}
   A_\np^{\textrm{QCD}}(\{p\})\right)\;\;,
\label{Eq004}		     
\end{align}
where the sum is over all permutations of $(1,\ldots,\np)$. 
An efficient 
algorithm to calculate a sum over permutations can be found in \cite{Kuijf}.

When doing calculations with this algorithm on a phase space cut such that
$(p_i+p_j)^2>s_0$ for all $i\neq j$ and some reasonable $s_0>0$, we notice that
a very high percentage of the generated events does not pass the cuts.  An
important reason why this happens is that the cuts, generated by the choices
of $g$ (\eqn{Eq002}) and $\xim$ (\eqn{Eq003}), are implemented only on
the variables $\xi^{i,j}_{k,l}$ that appear explicitly in the generation of the
QCD-antenna. Therefore, an improvement is obtained as follows.  Let $\Pol_m$
denote the subspace of $[-1,1]^m$ for which $|x_i-x_j|\leq1$ for all
$i,j=1,\ldots,m$, and let us denote the number of $\xi^{i,j}_{k,l}$-variables
that has to be generated $n_\xi \df 2\np-4$. An improvement is obtained if the
generation of these variables is replaced by 
\begin{Alg}[\texttt{IMPROVEMENT}]
\begin{enumerate}
\item generate $(x_1,\ldots,x_{\np_\xi})$ distributed uniformly in 
      $\Pol_{n_\xi}$;
\item define $x_0\df0$ and put, for all $i=2,\ldots,\np-1$,  
      \begin{equation}
         \xi^{i-1,i}_{i-1,\np}\lar e^{(x_{2i-3}-x_{2i-4})\log\xim}\;\;,\quad
	 \xi^{i,\np}_{i-1,\np}\lar e^{(x_{2i-2}-x_{2i-4})\log\xim}\;\;.
	 \label{Eq006}
      \end{equation}
\end{enumerate}
\end{Alg}
Because all the variables $x_i$ are distributed uniformly such that
$|x_i-x_j|\leq1$, {\em all} quotients $\xi^{i,j}_{k,l}$ with $(i,j)$ and
$(k,l)$ in $\{(i-1,i)\,,\,(i,\np)\,|\,i=2,\ldots,\np-1\}$ are distributed such
that they satisfy $\xim^{-1}\leq\xi^{i,j}_{k,l}\leq\xim$. This is an
improvement on the previous situation, because then only the quotients
$\xi^{i-1,1}_{i-1,\np}$ and $\xi^{i,\np}_{i-1,\np}$ with $i=2,\ldots,\np-1$
satisfied the relation. In terms of the variables $x_i$, this means that the
volume of $\Pol_{n_\xi}$ is generated, which is $n_\xi+1$, instead of the
volume of $[-1,1]^{n_\xi}$, which is $2^{n_\xi}$.  We have to note here that
this improvement only makes sense because there is a very efficient algorithm
to generate the uniform distribution in $\Pol_m$ \cite{HKpol}. The total
density changes such that the product of the $g$-functions in \eqn{Eq005} has
to be replaced by 
\begin{equation}
   g^{\Pol}_{\np-2}(\xim;\{\xi\}) 
   \;\df\; \frac{1}{(n_\xi+1)(\log\xim)^{n_\xi}}\times
           \begin{cases}
	      1 &\textrm{if $(x_1,\ldots,x_{n_\xi})\in\Pol_{n_\xi}$} \;,\\
	      0 &\textrm{if $(x_1,\ldots,x_{n_\xi})\not\in\Pol_{n_\xi}$}\;,
	   \end{cases}
\end{equation} 
where the variables $x_i$ are functions of the variables $\xi^{i,j}_{k,l}$ as
defined by $(\ref{Eq006})$.  Again, only the normalization has to be
calculated for the weight of an event.

We compare \sarge\ with \rambo\ in the calculation of the cross-section of the
processes 
\begin{equation}
   \epl\emi\longrightarrow\gamma^*\longrightarrow
   \q\qb\gl,\;\q\qb\q\qb,\;\q\qb\q'\qb',\;\q\qb\gl\gl,
   \;\q\qb\q\qb\gl,\;\q\qb\q'\qb'\gl,\;\q\qb\gl\gl\gl \;\;.
\end{equation}
The squared matrix element was calculated with the algorithm presented in
\cite{DKP}, suitably adapted for these processes. We used massless electrons and
quarks, and took the sum over final-state helicities and the average over
initial-state helicities.  We also summed over the color configurations of the
final states. The center-of-mass energy $\sqrt{s}$ was fixed to $500$ \GeV\ for
the processes with $5$ outgoing momenta, and to $100$ \GeV\ for the other
processes. The cuts on the phase space where fixed with choices of a parameter
$\tau$, which is related to the cut-off $s_0$ on the squares of the outgoing
momenta (\eqn{Eq007}) by 
\begin{equation}
   s_0 = \frac{2s\tau}{\np(\np-1)} \;\;,
\end{equation}
where $\np$ is the number of outgoing momenta.  If $\tau=1$, then $s_0$ is
larger than the maximal value that is kinematically allowed. The couplings and
charges in various processes were all set to the value $1$, since they only
contribute a factor to the cross-section, which is irrelevant for this
analysis. The results of the computer runs are given in the tables below.
Presented are the final result for the cross-section $\sigma$ in units of
$\GeV^{-2}$, the number of generated events $\Nge$, the number of accepted
events $\Nac$, and the cpu-time consumed $\tcpu$ in seconds. All Monte Carlo
runs were performed on a single 440-MHz UltraSPARC-IIi processor, and were
stopped when an expected error of $3\%$ was reached.

The final results for the cross-sections are irrelevant in our discussion, and
are just printed to show that the results with \sarge\ and \rambo\ are
compatible within the $3\%$ error estimate. The most important conclusion 
that can be drawn from the results is that \sarge\ needs less accepted 
events than \rambo\ for the given error estimate, especially 
for small values of $\tau$, {\em i.e.}, for phase space that comes close to 
the singularities of the QCD-amplitudes. (Remember that the ratio of 
the volumes of cut phase space and whole phase space is given by $\Nac/\Nge$ 
for \rambo.)
As a result, less evaluations of the 
matrix elements have to be done which accounts for a large gain in computer 
time. It is true that \sarge\ is ``ineffective'' in the sense that many of 
the generated events have to be rejected because they do not satisfy the 
cuts imposed, but this is fully compensated by the fact that generating 
random numbers is much cheaper than evaluating matrix elements nowadays.
For the last four processes, no results with \rambo\ and $\tau=0.01$ are 
presented, but we observe that $\tcpu>130,000$ seconds. The fraction of 
phase space covered with five massless momenta and $\tau=0.01$ is 
$0.893\pm0.001$.

\begin{center}
\begin{tabular}{|>{$}c<{$}||>{$}c<{$}|>{$}c<{$}||>{$}c<{$}|>{$}c<{$}|
                           |>{$}c<{$}|>{$}c<{$}||>{$}c<{$}|>{$}c<{$}|}
  \hline \multicolumn{9}{|>{$}c<{$}|}{\epl\emi\rightarrow\q\qb\gl}\\
  \hline \multicolumn{1}{|>{$}c<{$}||}{\tau }&
         \multicolumn{2}{>{$}c<{$}||}{0.5}&   
         \multicolumn{2}{>{$}c<{$}||}{0.1}&   
         \multicolumn{2}{>{$}c<{$}||}{0.05}&   
         \multicolumn{2}{>{$}c<{$}|}{0.01}\\   
  \hline \textrm{alg.} &\sarge\ &\rambo\ &\sarge\ &\rambo\ 
                       &\sarge\ &\rambo\ &\sarge\ &\rambo\ \\
  \hline \sigma &1.85\emu{5} &1.85\emu{5} &1.53\emu{4} &1.58\emu{4} 
&2.61\emu{4} &2.66\emu{4} &6.26\emu{4} &6.41\emu{4} \\
  \hline \Nge   &7,691 &25,782 &10,777 &24,801 &10,806 &37,121 &11,437 
&366,614 \\
  \hline \Nac   &5,503 &6,536 &9,436 &20,112 &9,852 &33,577 &10,860 &359,447 \\
  \hline \tcpu  &251 &293 &429 &899 &451&1,503 &497 &16,124 \\\hline
\end{tabular}
\end{center}

\begin{center}
\begin{tabular}{|>{$}c<{$}||>{$}c<{$}|>{$}c<{$}||>{$}c<{$}|>{$}c<{$}|
                           |>{$}c<{$}|>{$}c<{$}||>{$}c<{$}|>{$}c<{$}|}
  \hline \multicolumn{9}{|>{$}c<{$}|}{\epl\emi\rightarrow\q\qb\q\qb}\\
  \hline \multicolumn{1}{|>{$}c<{$}||}{\tau}&
         \multicolumn{2}{>{$}c<{$}||}{0.5}&   
         \multicolumn{2}{>{$}c<{$}||}{0.1}&   
         \multicolumn{2}{>{$}c<{$}||}{0.05}&   
         \multicolumn{2}{>{$}c<{$}|}{0.01}\\   
  \hline \textrm{alg.} &\sarge\ &\rambo\ &\sarge\ &\rambo\ 
                       &\sarge\ &\rambo\ &\sarge\ &\rambo\ \\
  \hline \sigma &9.79\emu{9} &10.4\emu{9} &7.72\emu{7} &7.86\emu{7} 
&1.90\emu{6} &1.83\emu{6} &7.39\emu{6} &7.00\emu{6} \\
  \hline \Nge   &64,384 &158,678 &32,492 &27,091 &34,701 &29,642 &41,744 
&113,368 \\
  \hline \Nac   &4,428 &4,551 &9,894 &15,328 &13,081 &22,297 &20,150 &107,021\\
  \hline \tcpu  &775 &786 &1,718 &2,606 &2,256 &3,778 &3,578 &18,038 \\\hline
\end{tabular}
\end{center}

\begin{center}
\begin{tabular}{|>{$}c<{$}||>{$}c<{$}|>{$}c<{$}||>{$}c<{$}|>{$}c<{$}|
                           |>{$}c<{$}|>{$}c<{$}||>{$}c<{$}|>{$}c<{$}|}
  \hline \multicolumn{9}{|>{$}c<{$}|}{\epl\emi\rightarrow\q\qb\q'\qb'}\\
  \hline \multicolumn{1}{|>{$}c<{$}||}{\tau}&
         \multicolumn{2}{>{$}c<{$}||}{0.5}&   
         \multicolumn{2}{>{$}c<{$}||}{0.1}&   
         \multicolumn{2}{>{$}c<{$}||}{0.05}&   
         \multicolumn{2}{>{$}c<{$}|}{0.01}\\   
  \hline \textrm{alg.} &\sarge\ &\rambo\ &\sarge\ &\rambo\ 
                       &\sarge\ &\rambo\ &\sarge\ &\rambo\ \\
  \hline \sigma &5.38\emu{9} &5.30\emu{9} &4.07\emu{7} &4.24\emu{7} 
&1.00\emu{6} &1.02\emu{6} &3.95\emu{6} &3.89\emu{6} \\
  \hline \Nge   &98,840 &245,138 &50,052 &45,963 &63,398 &50,873 &71,254 
&366,166 \\
  \hline \Nac   &6,696 &7,022 &15,392 &25,883 &23,989 &38,145 &34,584 &345,323 
\\
  \hline \tcpu  &1,165 &1,198 &2,664 &4,346 &4,133 &6,434 &5,843 &58,708 
\\\hline
\end{tabular}
\end{center}

\begin{center}
\begin{tabular}{|>{$}c<{$}||>{$}c<{$}|>{$}c<{$}||>{$}c<{$}|>{$}c<{$}|
                           |>{$}c<{$}|>{$}c<{$}||>{$}c<{$}|}
  \hline \multicolumn{8}{|>{$}c<{$}|}{\epl\emi\rightarrow\q\qb\gl\gl}\\
  \hline \multicolumn{1}{|>{$}c<{$}||}{\tau}&
         \multicolumn{2}{>{$}c<{$}||}{0.5}&   
         \multicolumn{2}{>{$}c<{$}||}{0.1}&   
         \multicolumn{2}{>{$}c<{$}||}{0.05}&   
         \multicolumn{1}{>{$}c<{$}|}{0.01}\\   
  \hline \textrm{alg.} &\sarge\ &\rambo\ &\sarge\ &\rambo\ 
                       &\sarge\ &\rambo\ &\sarge\  \\
  \hline \sigma &1.76\emu{7} &1.70\emu{7} &1.86\emu{5} &1.95\emu{5} 
&5.19\emu{5} &5.27\emu{5} &5.40\emu{4} \\
  \hline \Nge   &96,942 &268,407 &42,321 &86,608 &50,552 &298,073 &50,414 \\
  \hline \Nac   &6,579 &7,677 &12,945 &48,902 &19,091 &223,530 &26,551 \\
  \hline \tcpu  &1,363 &1,597 &3,619 &6,398 &3,802 &43,913 &5,287  
\\\hline
\end{tabular}
\end{center}

\begin{center}
\begin{tabular}{|>{$}c<{$}||>{$}c<{$}|>{$}c<{$}||>{$}c<{$}|>{$}c<{$}|
                           |>{$}c<{$}|>{$}c<{$}||>{$}c<{$}|}
  \hline \multicolumn{8}{|>{$}c<{$}|}{\epl\emi\rightarrow\q\qb\q\qb\gl}\\
  \hline \multicolumn{1}{|>{$}c<{$}||}{\tau}&
         \multicolumn{2}{>{$}c<{$}||}{0.5}&   
         \multicolumn{2}{>{$}c<{$}||}{0.1}&   
         \multicolumn{2}{>{$}c<{$}||}{0.05}&   
         \multicolumn{1}{>{$}c<{$}|}{0.01}\\   
  \hline \textrm{alg.} &\sarge\ &\rambo\ &\sarge\ &\rambo\ 
                       &\sarge\ &\rambo\ &\sarge\  \\
  \hline \sigma &2.04\emu{11} &1.91\emu{11} &4.05\emu{8} &4.08\emu{8} 
&1.68\emu{7} &1.61\emu{7} &1.48\emu{6} \\
  \hline \Nge   &4,028,648 &4,017,888 &238,220 &97,035 &203,237 &210,325 
&176,710 \\
  \hline \Nac   &5,616 &5,094 &14,216 &33,239 &19,522 &121,734 &29,492 \\
  \hline \tcpu  &4,530 &3,941 &10,333 &23,875 &14,159 &87,756 &21,407  
\\\hline
\end{tabular}
\end{center}

\begin{center}
\begin{tabular}{|>{$}c<{$}||>{$}c<{$}|>{$}c<{$}||>{$}c<{$}|>{$}c<{$}|
                           |>{$}c<{$}|>{$}c<{$}||>{$}c<{$}|}
  \hline \multicolumn{8}{|>{$}c<{$}|}{\epl\emi\rightarrow\q\qb\q'\qb'\gl}\\
  \hline \multicolumn{1}{|>{$}c<{$}||}{\tau}&
         \multicolumn{2}{>{$}c<{$}||}{0.5}&   
         \multicolumn{2}{>{$}c<{$}||}{0.1}&   
         \multicolumn{2}{>{$}c<{$}||}{0.05}&   
         \multicolumn{1}{>{$}c<{$}|}{0.01}\\   
  \hline \textrm{alg.} &\sarge\ &\rambo\ &\sarge\ &\rambo\ 
                       &\sarge\ &\rambo\ &\sarge\ \\
  \hline \sigma &1.05\emu{11} &1.05\emu{11} &2.19\emu{8} &2.23\emu{8} 
&9.07\emu{8} &8.86\emu{8} &7.85\emu{7} \\
  \hline \Nge   &5,596,725 &6,929,475 &436,225 &188,693 &377,384 &522,602 
&305,426 \\
  \hline \Nac   &7,730 &8,844 &26,154 &64,558 &36,042 &302,724 &51,044 \\
  \hline \tcpu  &5,882 &6,494 &17,595 &43,104 &24,764 &201,801 &34,700  
\\\hline
\end{tabular}
\end{center}

\begin{center}
\begin{tabular}{|>{$}c<{$}||>{$}c<{$}|>{$}c<{$}||>{$}c<{$}|>{$}c<{$}|
                           |>{$}c<{$}|>{$}c<{$}||>{$}c<{$}|}
  \hline \multicolumn{8}{|>{$}c<{$}|}{\epl\emi\rightarrow\q\qb\gl\gl\gl}\\
  \hline \multicolumn{1}{|>{$}c<{$}||}{\tau}&
         \multicolumn{2}{>{$}c<{$}||}{0.5}&   
         \multicolumn{2}{>{$}c<{$}||}{0.1}&   
         \multicolumn{2}{>{$}c<{$}||}{0.05}&   
         \multicolumn{1}{>{$}c<{$}|}{0.01}\\   
  \hline \textrm{alg.} &\sarge\ &\rambo\ &\sarge\ &\rambo\ 
                       &\sarge\ &\rambo\ &\sarge\  \\
  \hline \sigma &1.31\emu{11} &1.30\emu{11} &3.63\emu{7} &3.54\emu{7} 
&1.63\emu{6} &1.54\emu{6} &1.85\emu{5} \\
  \hline \Nge   &5,926,016 &6,650,538 &366,538 &131,617 &303,003 &186,257 
&335,307 \\
  \hline \Nac   &8,194 &8,475 &21,918 &45,157 &29,018 &107,897 &56,008 \\
  \hline \tcpu  &7,407 &7,398 &18,120 &36,958&24,036 &88,318 &46,673 \\\hline
\end{tabular}
\end{center}

As an extra illustration, we also present the convergence to zero of the 
expected error during the Monte Carlo-run for a few cases. In \fig{figqqg}, we 
plot the 
relative error as function of the number of generated events using a 
double-log scale. We first of all observe that the curves for \sarge\ are less 
spiky, which shows that \sarge\ takes care for a substantial part of the 
singular behavior of the integrand. Every time a \rambo-event hits a 
singularity, a term much larger than the average so far is added to the 
Monte Carlo sum, resulting in an increase of the expected error. 
Furthermore, we observe that the \sarge-error converges quicker than the 
\rambo-error, except in the case of $\epl\emi\ra\q\qb\gl\gl\gl$ with 
$\tau=0.05$. However, this is a plot of the error as function of the number 
of generated events, and we know that many \sarge-events have to be rejected. 
A more realistic view is given by a plot of the error as function of cpu-time
(\fig{figcpu}), which clearly shows that \sarge\ outperforms \rambo. 

\begin{figure}
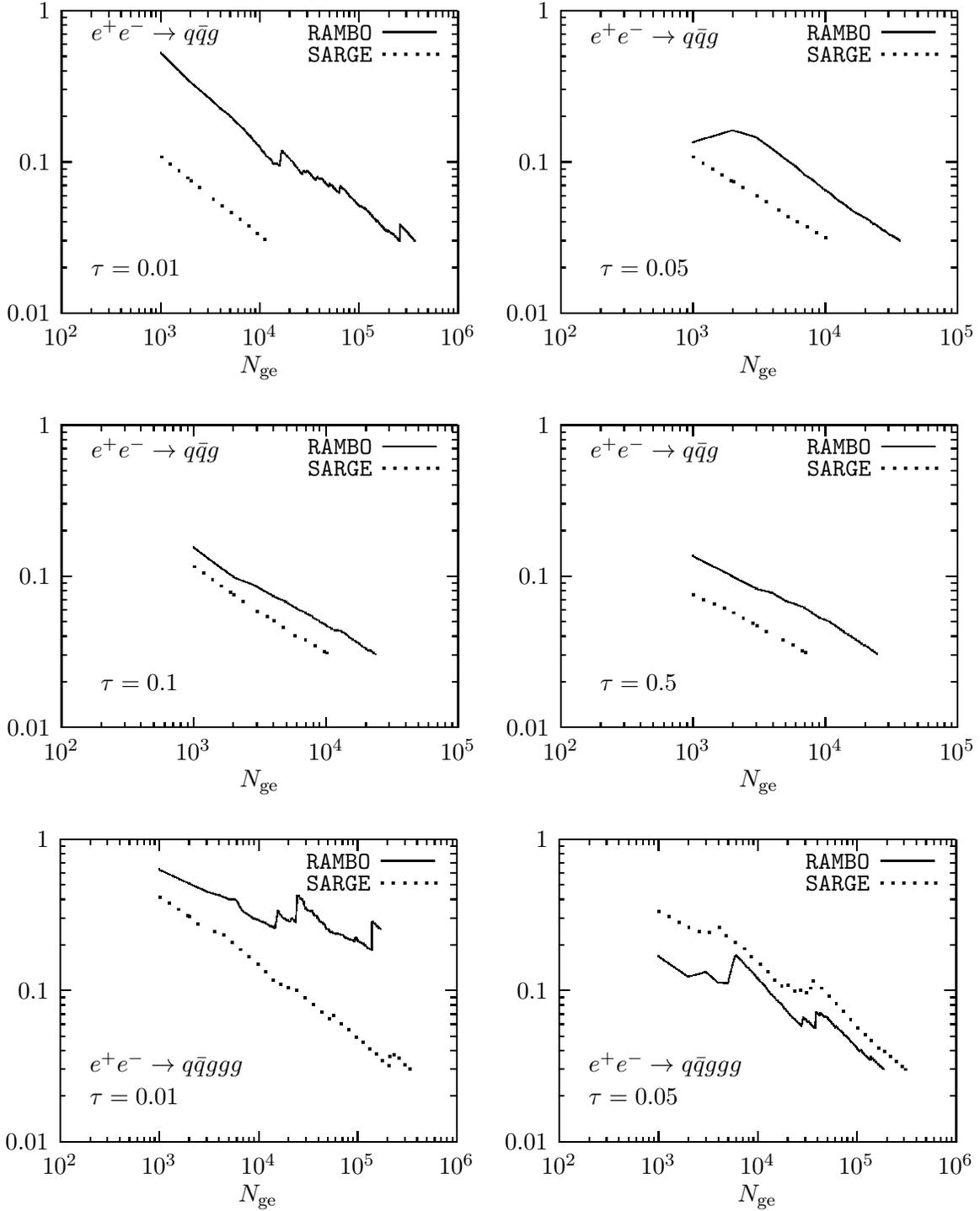

\begin{center}
\hspace{-20pt}
\vspace{-10pt}
%
%
%
\setlength{\unitlength}{0.240900pt}
\ifx\plotpoint\undefined\newsavebox{\plotpoint}\fi
\sbox{\plotpoint}{\rule[-0.200pt]{0.400pt}{0.400pt}}%

\end{center}
\caption{The expected relative error as function of the number of generated 
         events.}
\label{figqqg}
\end{figure}

\begin{figure}
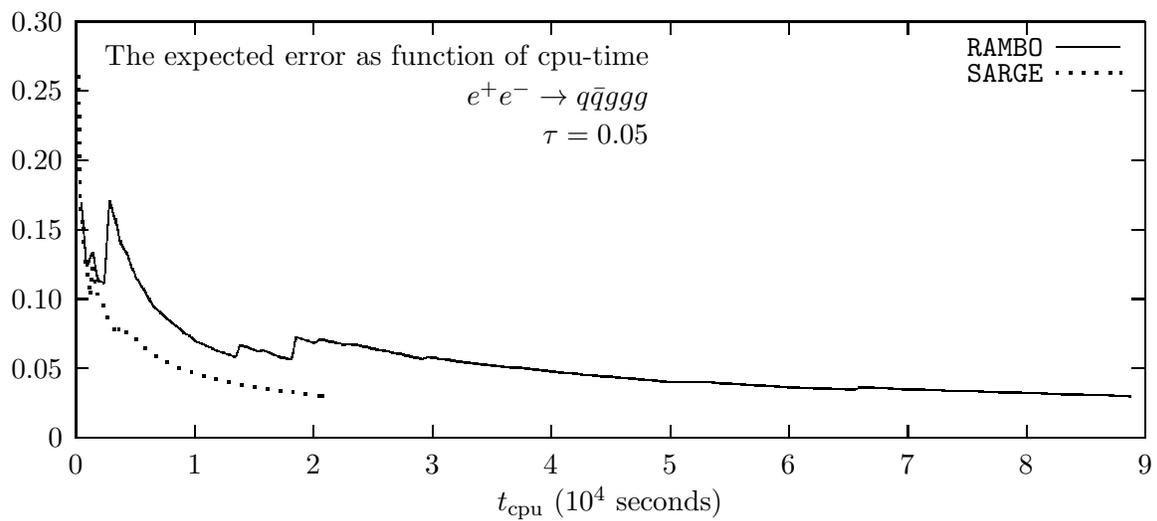

\begin{center}
%
%
\vspace{-10pt}
\setlength{\unitlength}{0.240900pt}
\ifx\plotpoint\undefined\newsavebox{\plotpoint}\fi
%
\caption{The expected relative error as function of cpu-time.}
\label{figcpu}
\end{center}
\end{figure}

\end{document}